\documentclass[prb,twocolumn,floatfix,preprintnumbers,amsmath,amssymb,superscriptaddress,showpacs]{revtex4-1}

\usepackage{graphicx}
\usepackage{graphics}
\usepackage{dcolumn}
\usepackage{bm}
\usepackage{xspace}
\usepackage{color}

\newcommand{\PHCX} {(C$_4$H$_{12}$N$_2$)Cu$_2$(Cl$_{1-x}$Br$_{x}$)$_6$ \xspace}

\newcommand{\PHCC} {(C$_4$H$_{12}$N$_2$)Cu$_2$Cl$_6$ \xspace}

\newcommand{\IPACX}{(CH$_3$)$_2$CHNH$_3$Cu(Cl$_{1-x}$Br$_x$)$_3$ \xspace}

\begin{document}
\DeclareGraphicsExtensions{.pdf,.png,.gif,.jpg}

\title{Field-concentration phase diagram of a quantum spin liquid with bond defects.}

 \author{D. H\"uvonen}
 \affiliation{Neutron Scattering and Magnetism, Laboratory for Solid State Physics, ETH Zurich, Switzerland.}

\author{G. Ballon}
 \affiliation{Laboratoire National des Champs Magnétiques Intenses - Toulouse, Tolouse, France.}
 \author{A. Zheludev}
 \homepage{http://www.neutron.ethz.ch/}
 \affiliation{Neutron Scattering and Magnetism, Laboratory for Solid State Physics, ETH Zurich, Switzerland.}
\date{\today}

\begin{abstract}
The magnetic susceptibility of the gapped quantum spin liquid compound \PHCC and its chemically disordered derivatives \PHCX are systematically studied in magnetic fields of up to 45~T, as a function of Br concentration. The corresponding field-temperature and field-concentration phase diagrams are determined. Measurements on the disorder-free parent compound are not fully consistent with previously published results by other authors [PRL{\bf 96}, 257203 (2006)]. The effect of Br/Cl substitution on the magnetic properties is superficially similar to that of finite temperature. However, important differences are identified and discussed with reference to the previously studied magnetic excitation spectra.
\end{abstract}

\pacs{64.70.Tg,61.05.C-,75.10.Kt,74.62.En}

\maketitle

\section{Introduction}
Despite considerable theoretical interest, systematic experimental studies of quantum phase transitions (QPT) in the presence of disorder are still lagging behind.
For the disorder-free case, excellent experimental realization of various QPTs have been found among field-induced transitions in quantum spin liquid materials. \cite{Oosawa1999,Chen2001,Paduan2004,Giamarchi2008,Coldea2010,Schmidiger2012} Some of the same compounds, with additional chemical disorder, have recently been tapped to study QPTs in the presence of randomness.\cite{Hong2010PRBRC,Yamada2011,Yu2012,Zheludev2013} Particular attention has been given to the so-called quantum Bose Glass (BG) problem. \cite{Giamarchi1987,Fisher1989} The BG is a zero-temperature disordered quantum phase that in 3-dimensional spin liquids, in applied magnetic fields, is expected to precede the ordered ``Bose-Einstein condensate of magnons'' (BEC) state.
To this date, an unambiguous identification of the BG remains elusive due to ongoing debates on the critical behaviour\cite{Weichman2007,Weichman2008,Priyadarshee2006,Yu2012} and possible side-effects of chemical disorder \cite{Zheludev2013}. In this context, we hereby extend our previous studies\cite{Huevonen2012,Huevonen2012-2} of the model spin liquid material \PHCX (PHCX), where chemical substitution on the halogen site leads to bond strength disorder. As a first step, we revisit the phase diagram of the parent compound \PHCC (PHCC), where we find marked inconsistencies with previously published work by other authors.\cite{Stone2006,Stone2007} We then turn to the effect of  disorder on the bulk magnetic properties of PHCX in applied magnetic fields, all the way up to the saturation field.
We systematically map out the field-concentration phase diagram, bringing a new dimension to the previously determined field-temperature diagram of PHCC.\cite{Stone2006,Stone2007}

\section{Experimental}
For the present study, single crystal PHCX samples were prepared as in our previous work,\cite{Huevonen2012,Huevonen2012-2} using the technique outlined in Ref.~\onlinecite{Yankova2012}. Samples with 0, 0.5, 1, 2.5, 3.5, 5, 7.5, 10 and 12.5\% of nominal Br content were synthesized. The quality of the crystals was checked using a Bruker AXS single crystal diffractometer equipped with a cooled Apex-II detector. The lattice parameters and the precise Cl/Br site occupancies were refined with the SHELXTL program using in average over 5000 reflections.

\begin{figure}[tb]
\includegraphics[width=\columnwidth]{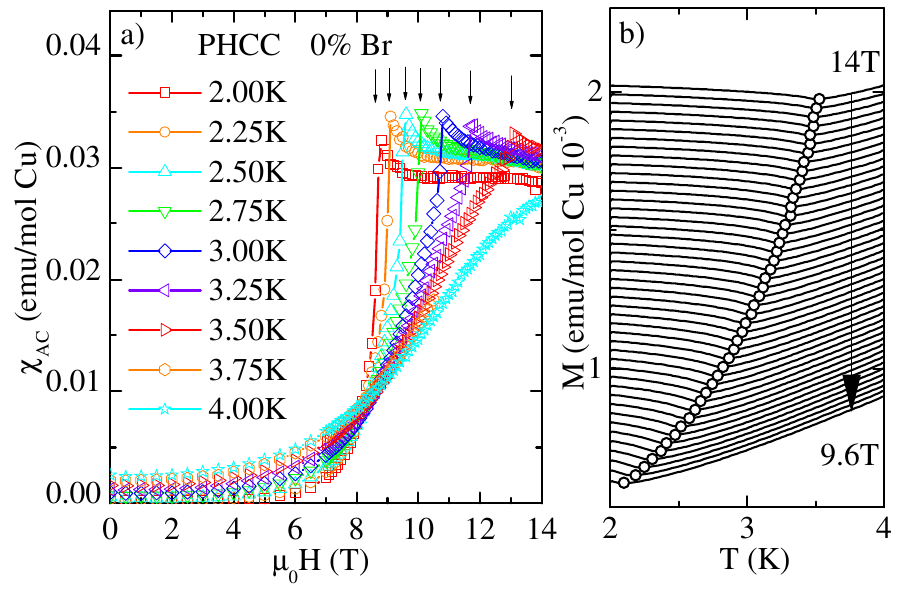}\\
\includegraphics[width=\columnwidth]{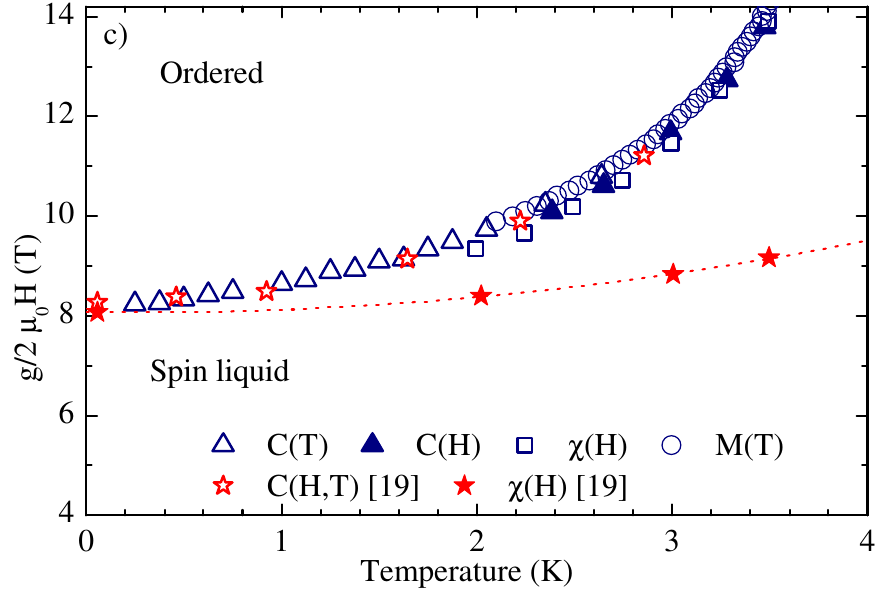}
\caption{(color online) a) AC susceptibility, $\chi(H)$ of PHCC at fields up to $14$~T from $T=2$~K to $4$~K, measured in $\mathrm{H}||\mathrm{c}$ orientation. b) Temperature dependence of magnetization, $M(T)$, at constant fields. Black hollow circles indicate positions of $M(T)$ minima. c) Measured $H-T$ phase diagram of PHCC. Symbols correspond to peaks in $C(H)$ (hollow triangles),\cite{Huevonen2012} $C(T)$ (filled triangles),\cite{Huevonen2012} $\chi(H)$ (squares) and dips in M(T) (circles). Data from Ref. \onlinecite{Stone2007} is shown by hollow and filled star symbols corresponding to specific heat and susceptibility anomalies, respectively. The dashed line is a guide to the eye.}
\label{acvsm}
\end{figure}

Bulk magnetic measurements up 14~T were carried out using the Quantum Designs Physical Property Measurement System. AC magnetic susceptibility was measured at 316 Hz. Although no frequency dependence up to 10kHz was detected in any of our samples, a low measurement frequency was used to avoid excessive sample heating at T=2K.
Sample masses ranged from 70 to 200 mg. All crystals used were well faceted and of an elongated shape along $c$-axis. The magnetic field was applied along that direction. The temperature dependence of magnetization at constant fields was measured using a vibrating sample magnetometer (VSM). High pulsed field magnetization measurements were conducted at Toulouse High Magnetic Field Facility using pulsed magnet with maximum field of 53~T and a $^4$He flow cryostat.
In this case, the samples consisted of 3-4 single crystals that were stacked along the magnetic field on the natural flat surfaces perpendicular to $\mathrm{a^*}$ axis. The orientation of each sample was checked with X-ray diffraction.

\section{Results and discussion}
\subsection{Magnetic properties of the parent compound.}
Before addressing the role of disorder, we revisit the phase diagram of the parent material. The field dependencies of magnetic susceptibility of PHCC, $\chi(H)=\mathrm{d}M/\mathrm{d}H$, measured at temperatures ranging between 2 to 4K, are shown in Fig.\ref{acvsm}a. The rapid increase of susceptibility at high fields can be associated with a Zeeman closure of the spin gap. At elevated temperatures, this increase becomes progressively broadened, while appearing almost step-like at low temperatures. This behavior is fully consistent with the results of Ref.~\onlinecite{Stone2001} and our expectations of a thermal broadening of the field-softened magnon excitations, due to collisions  at a finite temperature.\cite{Quintero2012} Despite the overall thermal broadening, at all temperatures, there is a cusp-like feature that remains sharp. Its temperature dependence is plotted in squares in Fig.\ref{acvsm}c. A well-known signature of field-induce transverse long-range ordering (BEC of magnons) is a characteristic dip in the temperature dependence of magnetization measured at a constant field.\cite{Nikuni2000} This dip is clearly seen in our measurements on PHCC as well, as shown in Fig.\ref{acvsm}b. In Fig.\ref{acvsm}c the temperature dependence of this anomaly is shown in circles.

From Fig.~\ref{acvsm}c, we see that both the $M(T)$ anomaly and the susceptibility cusp occur simultaneously. Moreover, they coincide with the previously studied $\lambda$-anomaly in magnetic susceptibility of Refs.~\onlinecite{Stone2007,Huevonen2012} (triangles and open stars). To make this comparison between experiments performed in different geometries possible, in Fig.~\ref{acvsm}c we have normalized the magnetic field by the known projections of the anisotropic $g$-factor ($g_b=2.2,g_{\perp b}=2.05,g_c=2.12$).\cite{Glazkov2012} The clear conclusion is that all three features correspond to the onset of 3-dimensional long range order, and are to be associated with the magnetic BEC transition.

These results are in contradiction with previous studies of Refs.~\onlinecite{Stone2006,Stone2007} by Stone {\it et al.}. There it was found that a jump in magnetic susceptibility broadens only slightly with temperature, certainly much less than in our measurements (for a direct comparison see Fig.~2 in Ref.~\onlinecite{Stone2006}). Interestingly, earlier data by the same authors, which unfortunately only extend up to $T=1.8$~K, showed exactly the same kind of progressive thermal broadening as we observe (see Fig.~3 in Ref.~\onlinecite{Stone2001}). According to Refs.~\onlinecite{Stone2006,Stone2007}, the susceptibility jump is at a lower field than the specific heat anomaly, and that the difference increased with temperature. For a comparison with our findings, the temperature of the susceptibility jump digitized from Ref.~\onlinecite{Stone2007} is plotted in solid stars in Fig.\ref{acvsm}c.
In Ref.~\onlinecite{Stone2007}, the intermediate $H-T$ region of non-zero susceptibility (above the susceptibility jump) and no long range order (below the specific heat anomaly) was termed ``renormalized classical phase''. It was speculated that there a small number of excitations behave as free particles in the two-dimensional magnetic planes. Clearly, no signatures of such a phase are present in our measurements, where the susceptibility jump always coincides with the specific heat divergence. The cause of the discrepancy can not be unambiguously identified. However, the almost complete lack of a thermal broadening of $\chi(H)$ curves in Ref.\onlinecite{Stone2006}  may indicate a sample thermalization problem in the pulsed field experiment of Stone {\it et al}.

\subsection{Structural effects of Br substitution}
\begin{figure}[tb]
\includegraphics[width=9cm]{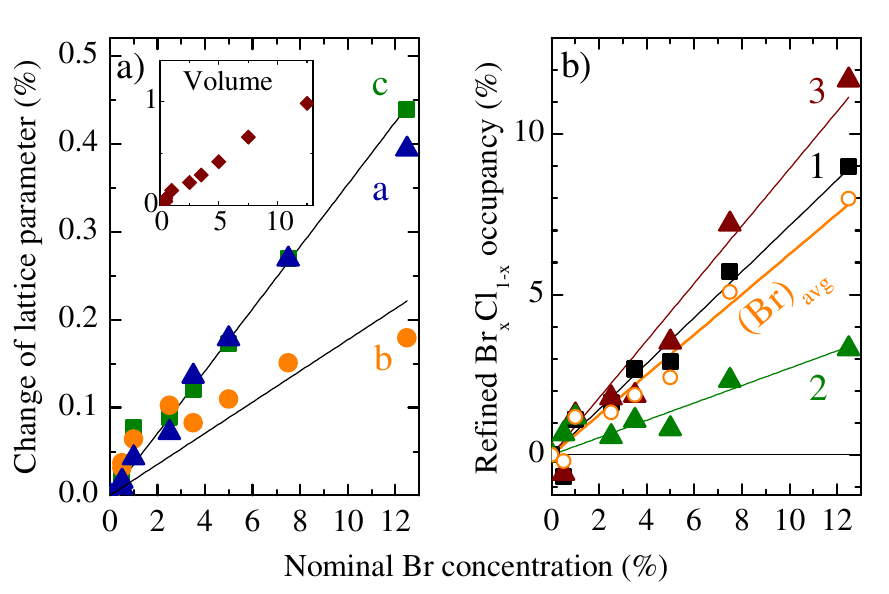} \\
\includegraphics[width=8cm]{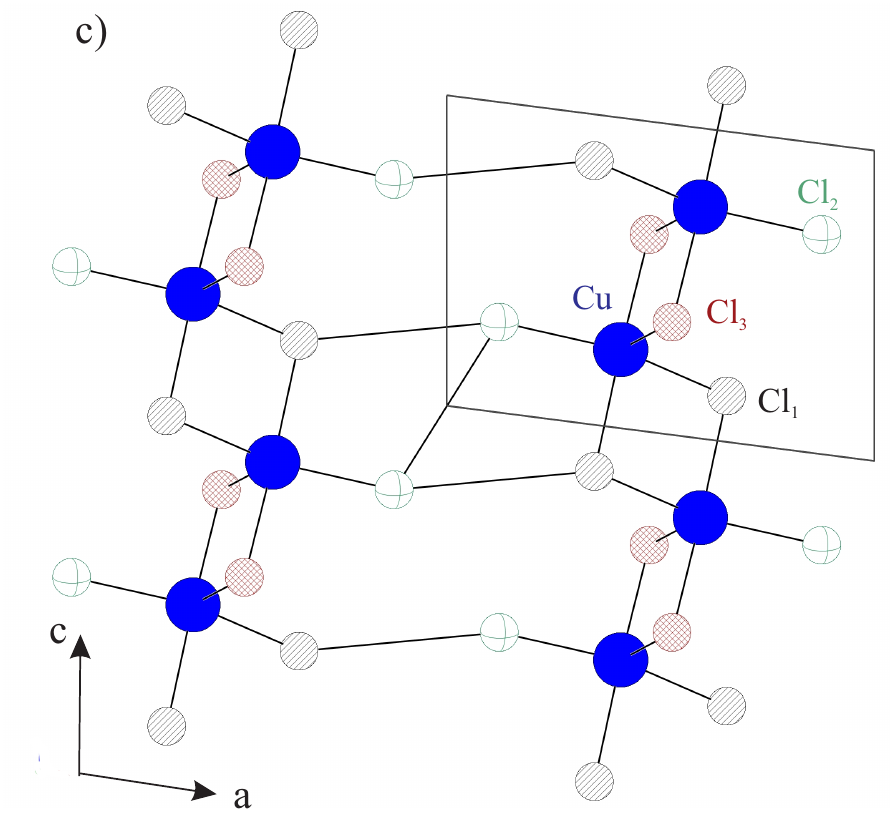}
\caption{(color online) Relative increase of lattice parameters and unit cell volume (a), and   Br/Cl occupancies at the three unique positions in unit cell (b), as a function of nominal Br concentration in PHCX. Thin lines are guide to the eye and indexes refer to Cl position as shown in (c). The solid orange line is a linear fit to the average Br occupancy at Cl sites shown by orange circles. 
(c): A projection of the Cu$_2$Cl$_6$ layers onto a plane perpendicular to the  crystallographic $b$ axis in the structure of PHCC. Blue filled circles are the magnetic Cu$^{2+}$ ions. Shaded circles correspond to the 3 crystallographically unique halide ion positions.}
\label{xrd}
\end{figure}

The first step in understanding the role of Br substitution is to identify its precise influence on the crystal structure of \PHCX. The most obvious effect is a systematic increase of lattice constants and unit cell volume with increasing nominal Br content $x$, as shown in  Fig.\ref{xrd}a, which is based on the results of single crystal X-ray diffraction measurements. Interestingly, the largest effect is on the $a$ and $c$ lattice constants. This is consistent with the layered structure of PHCC,\cite{Stone2001} where the magnetic ions and superechange-carrying halide ions form well-connected networks in the $(a,c)$ plane and are loosely stacked along the $b$ axis.

Our X-ray diffraction data were complete enough to also determine the local Cl/Br occupancies on each of the three crystallographically inequivalent halogen sites. The three refined Br occupancies are plotted vs. nominal Br content in the starting crystal growth solution in Fig.~\ref{xrd}b. The immediate conclusion is that Br ions primarily affect the nearest-neighbor and next-nearest neighbor superexchange pathways along the $c$ axis of PHCC (Fig.~\ref{xrd}c). In the nomenclature of Ref.~\onlinecite{Stone2001}, it is the exchange constants $J_1$ and $J_2$, respectively. In contrast, The actual Br insertion rates into the superexchange bonds along that $a$ axis are less than half of the nominal.  Averaging the three population factors we find that in the studied concentration range the actual Br content in crystalline samples is related to the nominal one in the growth solution through $x_\text{actual} = A\cdot x_\text{nominal}$, with $A=0.63(3)$.

\subsection{Magnetic properties of PHCX in fields up to 14~T}
\begin{figure}[tb]
\includegraphics[width=\columnwidth]{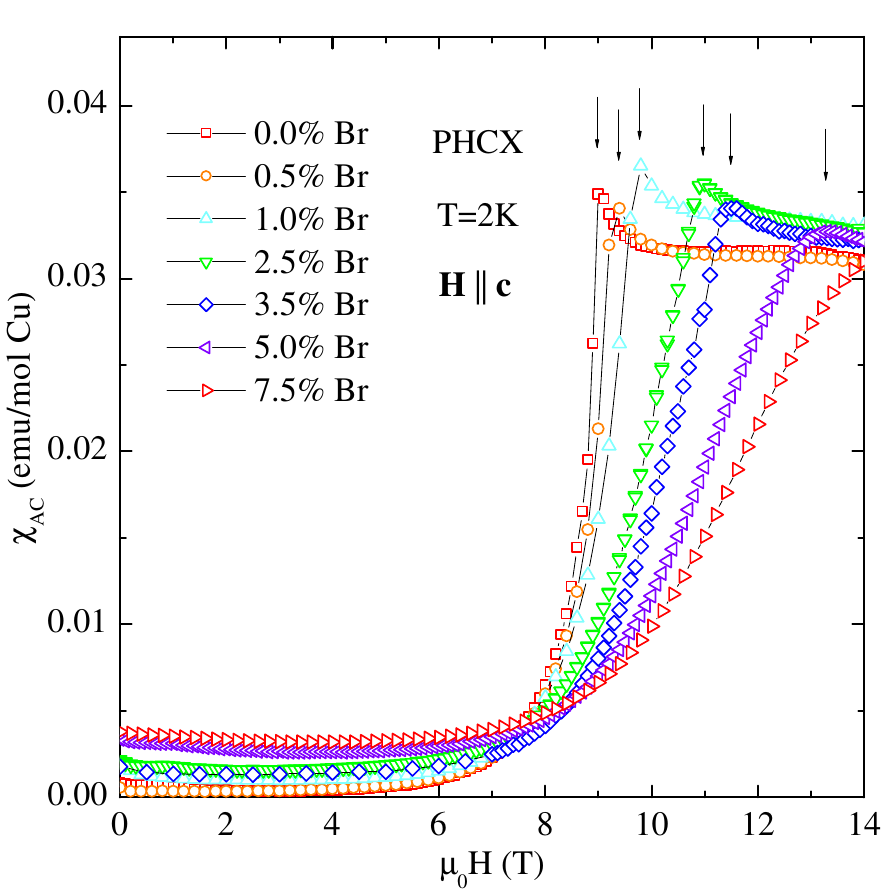}
\caption{(color online) AC susceptibility of PHCX at T=2K for 0 to 7.5\% Br content up to 14T applied in $\mathrm{H}||\mathrm{c}$ orientation. Arrows indicate the local susceptibility maxima.}
\label{acbr}
\end{figure}

The field dependencies of magnetic susceptibility measured in a series of \PHCX samples with varying Br content at $T=2$~K are shown in Fig.\ref{acbr}. At a first glance, the overall trend of increased disorder appears similar to that of finite temperature: the susceptibility jump broadens and moves to higher fields. The overall upward shift is clearly a consequence of the increase of the spin gap in Br-substituted samples.\cite{Huevonen2012-2} The broadening, on the other hand, is easy to understand in the framework of BG physics. As discussed in Refs.~\onlinecite{Fisher1989,Yu2012,Zheludev2013} and references therein, the appearance of non-zero uniform magnetic susceptibility below the critical field of 3-dimensional ordering is one of the key signatures of the BG. It represents a finite compressibility of the bosonic matter in the absence of BEC. By analogy with previous studies on \IPACX,\cite{Hong2010PRBRC} in our case of PHCX, this regime is to be associated with the field range where $\chi(H)$ increases steadily from a small value, just below the transition field, i.e., $H\gtrsim 8$~T.

\begin{figure}[tb]
\includegraphics[width=\columnwidth]{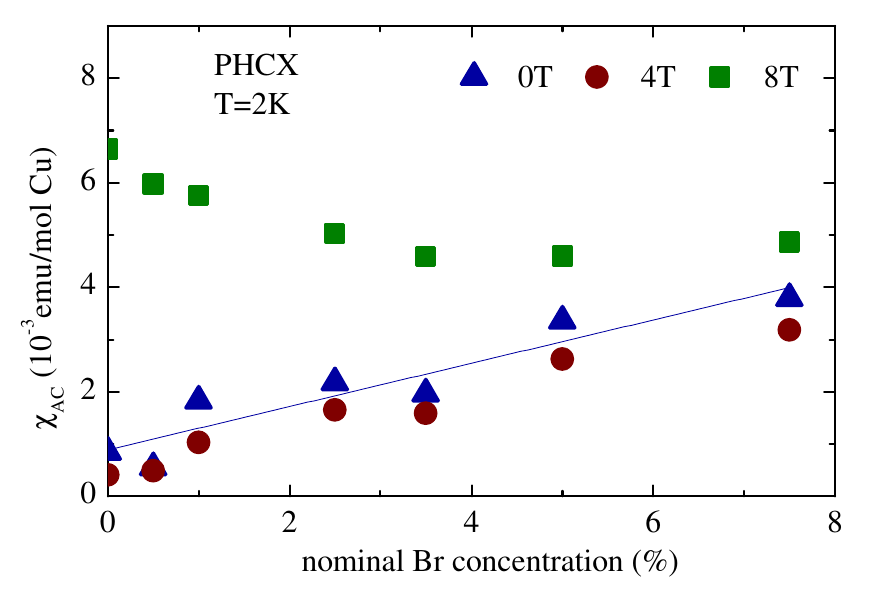}
\caption{(color online) Magnetic susceptibility at fields below the critical field of 3-dimensional ordering. Blue triangles, red circles and green squares correspond to AC susceptibility values shown in Fig.\ref{acbr} at 0, 4 and 8T fields, respectively. The solid line is a guide for the eye.}
\label{chiath}
\end{figure}

A more careful look at the data reveals important additional peculiarities.
First, we note that the sharp susceptibility cusp that in the pure sample we identified with 3-dimensional ordering becomes progressively broadened when disorder is introduced in PHCX. This is in contrast to the finite-$T$ effect on PHCC discussed above, where the kink remains sharp at all temperatures.
A second very interesting feature is that in all disordered samples, at low temperature, we observe additional magnetic susceptibility {\it inside} the gapped phase (i.e., below the critical field).
As shown in Fig.~\ref{chiath} (circles), the zero-field susceptibility increases monotonically with Br content. This contribution is {\it not} due to paramagnetic impurities. Indeed, in a magnetic field of 4~T (triangles) it remains practically unchanged. At $T=2$~K, under these conditions, paramagnetic $S=1/2$ impurities would be already close to saturation, with the susceptibility reduced by a factor of 4 compared to zero field. We conclude that the observed extra susceptibility below 4~T is of a Van Vleck character. At still higher fields, the trend is reversed in that the susceptibility decreases with increased Br concentration  (Fig.~\ref{chiath}, squares for $H=8$~T). This is clearly due to the contribution of the susceptibility jump at the critical field, which simply moves out to higher field with increased Br content.

It appears that the Van Vleck contribution at lower fields has a different origin, of which we can only speculate about at this stage. The simplest explanation is that each interstitial Br  distorts the local ionic environments of the Cu$^{2+}$ ions, affecting the gyromagnetic ratios of nearby Cu$^{2+}$ ions. In this situation, applying a uniform magnetic field will result in a non-uniform spin field. While the uniform magnetic susceptibility of an isotropic spin liquid is zero, the non-uniform one is not. Just as an illustration, consider the toy model of a simple $S=1/2$ dimer with a singlet-triplet gap $\Delta$, where the gyromagnetic factors $g_1$ and $g_2$ are different for the two spins. An external uniform magnetic field $H$ will induce both a uniform and a staggered component of the spin field, $h_\mathrm{u}=\mu_\mathrm{B}H(g_1+g_2)/2$ and $h_\mathrm{s}=\mu_\mathrm{B}H(g_1-g_2)/2$, respectively. At $T\rightarrow 0$, the susceptibility of the antiferromagnetic dimer to the first field is zero since a uniform field conserves total spin. However, the staggered spin susceptibility is non-zero and equal to $1/\Delta$. As a result, the net uniform magnetization will be $\mu_B^2H/\Delta(g_1-g_2)^2$.  The corresponding {\it uniform} magnetic susceptibility will be Van Vleck like in that it will be temperature and field independent for $g\mu_\mathrm{B}H, k_\mathrm{B}T\lesssim \Delta$. A similar effect will most certainly be present in PHCX, even though the exact nature of the spin liquid ground state is not known due to multiple exchange interactions.\cite{Stone2001} This mechanism is due to the local, ``differential'' influence of Br ions, so it makes perfect sense that the resulting Van Vleck term scales roughly linearly with Br content.

\subsection{Phase diagram in fields up to 45~T}
\begin{figure}[tb]
\includegraphics[width=8cm]{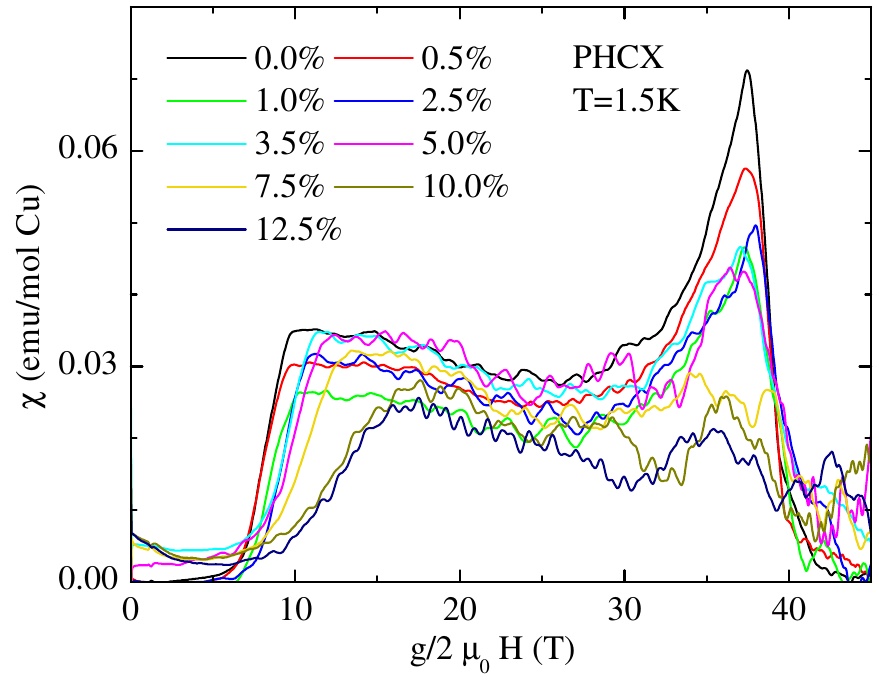}
\caption{(color online) Magnetic susceptibility of PHCX measured at 1.5~K in pulsed fields for several Br concentrations. The magnetic field was applied in $\mathrm{H}||\mathrm{a^*}$ orientation.}
\label{hfchi}
\end{figure}

One of the main results of this study are measurements of the magnetization of PHCX in pulsed magnetic fields, all the way up to the saturation field.
For a discussion of the results, it is more useful to focus on the differential susceptibility $\chi=\mathrm{dM}/\mathrm{dH}$, derived from the as-measured magnetization curves. Representative data collected at $T=1.5$~K are shown in  Fig.~\ref{hfchi}. The result for parent compound are in good qualitative agreement with previous measurements by Stone {\it et al.}.\cite{Stone2007}
For all Br concentrations, one clearly sees the typical top-hat shape, that spans across the ``compressible'' BEC phase.

\begin{figure}[tb]
\includegraphics[width=\columnwidth]{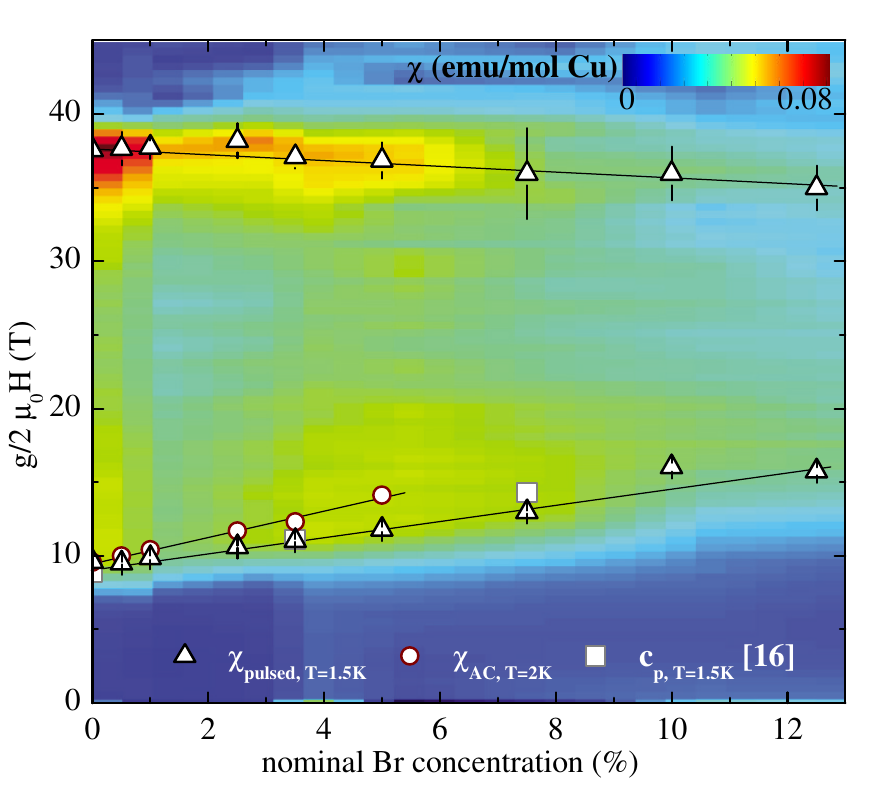}
\caption{(color online) False color map of magnetic susceptibility of PHCX plotted vs. magnetic field and nominal Br content. Black triangles and red circles indicate positions of susceptibility maxima in pulsed field and AC measurements, respectively. Gray squares indicate the critical fields obtained from specific heat measurements in Ref.\onlinecite{Huevonen2012}. Lines are guide to the eye.}
\label{hfchimap}
\end{figure}

The inherent noise in the data, primarily due to the small sample sizes, precludes a quantitative analysis. Despite that, for all concentrations it is still possible to roughly determine the upper and lower critical fields in order to establish a phase diagram. For these estimates, we took the intersection of linearly extrapolated data on either side of each of the two transitions. The thus obtained critical fields are plotted versus nominal Br content in  Fig.~\ref{hfchimap} (triangles), laid over a false color plot of the measured susceptibilities. For comparison, we also show the critical fields determined from the cusp in AC susceptibility at 2\,K and from the positions of specific heat anomalies at T=1.5~K.\cite{Huevonen2012}

The thus obtained field-concentration phase diagram is reminiscent of the field-temperature phase diagram determined in Ref.~\onlinecite{Stone2007}.
With increasing Br concentration the domain of the BEC phase shrinks, with the upper and lower critical fields decreasing and increasing, respectively. This behavior is fully consistent with the previously observed reduction of magnon bandwidth with Br substitution.\cite{Huevonen2012-2} At least two mechanisms could be liable for this behavior. First, disorder itself will limit the coherent propagation of quasiparticles and thereby reduce the bandwidth. In its nature, this effect is very similar to the reduction of bandwidth seen in spin liquids at a finite temperature due to magnon-magnon collisions (see, for example, Ref.~\onlinecite{Sasago1997}).

An alternative explanation is not related to disorder at all, but to the mean ``chemical pressure'' produced by Br interstitials. This pressure is manifested in the expansion of the crystal lattice and in this sense is {\it negative}, and is linear with Br concentration (see Fig.~\ref{xrd}). Direct hydrostatic studies of PHCC have demonstrated a linear decrease of the gap energy with applied (positive) pressure.\cite{Hong2010-2} If we interpret the concentration dependence of the critical fields in terms of Br-induced chemical pressure, we then naturally explain the linear behavior of the phase boundaries on Fig.~\ref{hfchimap}.  Note that these linear boundaries are the main distinction with the field-temperature phase diagram of the parent compound. There, both boundaries are curves, roughly corresponding to the BEC crossover critical exponent $\phi=2/3$.\cite{Stone2006,Huevonen2012}

\section{Conclusion}
We have performed a systematic study of bulk magnetic properties of the gapped quantum magnet PHCC and it's chemically disordered derivatives PHCX.
Contrary to previous reports,\cite{Stone2006} for PHCC, our experiments show no evidence of any crossover phase intermediate to spin liquid and long range ordered states. In PHCX the ordered phase shrinks roughly linearly with Br concentration.  In addition, the  transition to the magnetized phase becomes progressively broadened with increased disorder, not inconsistent with expectations for the BG regime. Finally, we detect an additional disorder-induced Van Vleck like contribution to low-field susceptibility that appears not to be directly related to the field-induced transition.

\section{Acknowledgements}
This work is partially supported by the Swiss National Fund under
project 2-77060-11 and through Project 6 of MANEP.
Work in high field facility was supported by Euromagnet II via the EU under Contract No. RII3-CT-2004-506239.
We thank Mr. S. Zhao, Dr. T. Yankova and Dr. V. Glazkov for their involvement in the early stages of this project.

\end{document}